# Security Logs to ATT&CK Insights: Leveraging LLMs for High-Level Threat Understanding and Cognitive Trait Inference

**Soham Hans[1], Stacy Marsella[2], Sofia Hirschmann[2], and Nikolos Gurney[1]**

[1]USC Institute for Creative Technologies, Playa Vista, CA 90094, USA
[2]Northeastern University, Boston, MA 02115, USA

**ABSTRACT**

Understanding adversarial behavior in cybersecurity has traditionally relied on high-level intelligence reports and manual interpretation of attack chains. However, real-time defense requires the ability to infer attacker intent and cognitive strategy directly from low-level system telemetry such as intrusion detection system (IDS) logs. In this paper, we propose a novel framework that leverages large language models (LLMs) to analyze Suricata IDS logs and infer attacker actions in terms of MITRE ATT&CK techniques. Our approach is grounded in the hypothesis that attacker behavior reflects underlying cognitive biases such as loss aversion, risk tolerance, or goal persistence that can be extracted and modeled through careful observation of log sequences. This lays the groundwork for future work on behaviorally adaptive cyber defense and cognitive trait inference. We develop a strategy-driven prompt system to segment large amounts of network logs data into distinct behavioral phases in a highly efficient manner, enabling the LLM to associate each phase with likely techniques and underlying cognitive motives. By mapping network-layer events to high-level attacker strategies, our method reveals how behavioral signals such as tool switching, protocol transitions, or pivot patterns correspond to psychologically meaningful decision points. The results demonstrate that LLMs can bridge the semantic gap between packet-level logs and strategic intent, offering a pathway toward cognitive-adaptive cyber defense.

## INTRODUCTION

Cyber defense operations rely heavily on interpreting large volumes of network telemetry, such as alerts and packet logs, to identify and respond to adversarial activity. However, traditional intrusion detection systems (IDS) like Suricata (The Open Information Security Foundation, 2025) produce fragmented and low-level event streams that capture "what happened" without conveying the meaning behind how they are connected. Bridging this semantic gap between packet-level indicators and high-level attacker intent remains a central challenge in cyber threat analysis. Current analytic pipelines depend on human expertise to correlate alerts, hypothesize MITRE ATT&CK (MITRE Corporation, 2025) techniques, and reconstruct adversarial strategies, a process that is time-intensive, error-prone, and cognitively demanding.

To address this limitation, we propose a structured large language model (LLM) (OpenAI, 2024) framework that directly infers attacker actions, phases, and cognitive tendencies from IDS logs. Our approach introduces a system that segments continuous network activity into meaningful behavioral phases and maps each phase to probable ATT&CK techniques supported by textual and temporal evidence. Beyond technical mapping, we explore how patterns such as tool switching, protocol transitions, and pivot behaviors can serve as signals of cognitive traits embedded in attacker decision-making. This builds on recent work demonstrating that cognitive biases such as loss aversion and ambiguity aversion manifest in hacker behavior during realistic cyber exercises (Beltz et al., 2025; Ferguson-Walter et al., 2018; Hitaj et al., 2025). By enabling LLMs to interpret low-level telemetry in strategic and psychological terms, this work moves toward a new paradigm of cognitive-adaptive cyber defense, where systems not only detect threats but also infer the human reasoning behind them.

## BACKGROUND AND RELATED WORK

The MITRE ATT&CK framework serves as a behavioral ontology for describing adversarial tactics, techniques, and procedures (TTPs). Recent surveys (Al-Sada et al., 2025; Roy et al., 2023) have highlighted ATT&CK's growing role in aligning observed activity with behavioral models in academic and behavioral settings. It provides a shared vocabulary for analysts to interpret observed activity, yet most operational uses depend on static signatures or manually defined alert-to-technique mappings. Such rule-based approaches often fail to capture contextual dependencies across hosts or time, limiting their ability to infer attacker strategy or intent from raw telemetry. Bridging this semantic gap between low-level alerts and high-level behavioral reasoning remains a central challenge in automated cyber defense.

Recent advances in large language models (LLMs) have shown promise for interpreting complex, unstructured data and inferring intent from behavioral narratives (Bunt et al., 2025; Peters & Matz, 2024). In cybersecurity research, LLMs have been applied primarily to human-authored artifacts such as analyst reports or attacker notes (Hans et al., 2025), where linguistic structure directly encodes reasoning. Extending these methods to machine-generated telemetry, such as Suricata IDS logs, introduces new difficulties: the absence of natural language cues, fragmented temporal context, and ambiguous behavioral boundaries. Addressing these issues requires structured prompting and segmentation strategies that expose latent behavioral meaning in raw event sequences.

Prior work (Hans et al., 2025) has explored using LLMs to infer attacker cognition and bias from structured human-authored records, demonstrating that traits such as loss aversion can be extracted from written accounts of hacker behavior. However, such data are rarely available in operational settings as defenders do not receive attacker notes, only machine logs. Building on that foundation, the present study extends cognitive inference into the domain of raw network telemetry, where behavioral patterns must be reconstructed from event sequences rather than explicit reasoning. By applying LLM-driven behavioral segmentation and ATT&CK-based mapping to Suricata IDS logs, this work advances the concept of cognitive-adaptive defense from controlled, note-based analysis toward real-time, data-driven interpretation of adversarial behavior.

## DATA: SURICATA-CENTRIC CORPUS CONSTRUCTION

The dataset used in this study originates from *Operation 418*, a controlled cybersecurity experiment designed to examine attacker decision-making in a realistic, adversarial environment. It is derived from the experiment described in (Beltz et al., 2025). In this setting, trained cybersecurity professionals acting as red-team participants were tasked with compromising a simulated enterprise network over a two-day engagement. The environment mirrored a mid-sized corporate infrastructure, featuring segmented internal subnets, domain controllers, web and file servers, and active blue-team monitoring through multiple security sensors. Each participant operated from an isolated virtual machine under restricted network access and limited external resources, simulating real-world operational constraints. As in prior analyses from this experiment, all participants provided informed consent, and data collection followed approved ethical protocols with anonymization of all identifiers.

In contrast to prior work that relied on participant-authored operational notes (OPNOTES), this study focuses exclusively on the machine telemetry captured during Operation 418. The primary source is the Suricata Intrusion Detection System (IDS), configured to log output encompassing alerts, flow records, and protocol metadata. Suricata served as the network-level observer, passively monitoring ingress and egress traffic across network segments. Each event record included timestamps, network identifiers such as source and destination addresses and ports, signature information, rule category, and summarized payload attributes. Because all traffic was captured within an encrypted tunnel, payload contents were not directly visible; therefore, analysis relied on flow-level characteristics (e.g., duration, packet counts, byte ratios) and metadata fields (e.g., TLS fingerprints, SNI values, certificate attributes) to reconstruct activity patterns.

This Suricata-centric corpus forms the foundation for our LLM-based behavioral analysis, enabling reasoning over coherent, interpretable units of attacker activity even when the underlying packet payloads remain encrypted.

## METHODOLOGY

### Overview

Our framework transforms low-level Suricata telemetry into interpretable behavioral representations through a two-stage process: (1) **action segmentation**, where related logs are grouped into discrete operator actions, and (2) **ATT&CK mapping**, where these actions are linked to higher-level adversarial techniques using a retrieval-augmented large language model (RAG-LLM). This hierarchical structure enables reasoning about attacker behavior at multiple levels of abstraction from packet-level evidence to cognitive strategy.

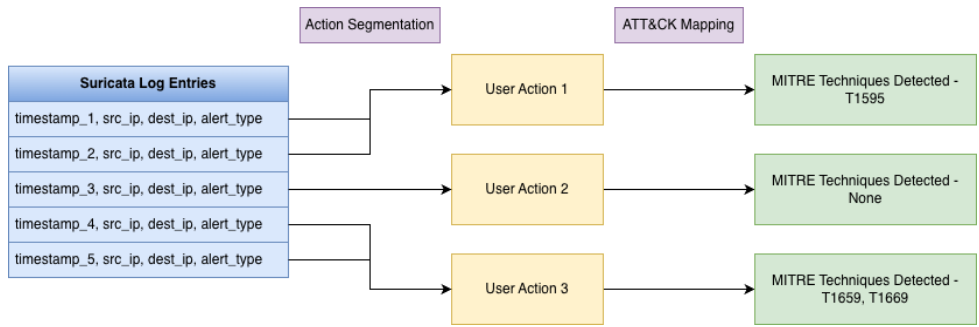

Figure 1. Framework Overview

## Action Segmentation from Suricata Logs

Suricata often produces multiple correlated alerts for a single user operation, leading to redundancy and fragmentation in raw telemetry. To recover meaningful behavioral units, we employ a large language model to iteratively segment the log stream into semantically coherent action groups. Each new log entry is evaluated in the context of a running summary representing the current group. The model determines whether the event continues the current action or marks the beginning of a new one. If continuing, the summary is updated; if new, the prior action summary is finalized, and a new grouping begins. This process produces a chronological sequence of concise, human-interpretable action descriptions, each corresponding to an underlying operator decision or system interaction.

## MITRE ATT&CK Mapping via Retrieval-Augmented Generation

Once the log stream is condensed into atomic actions, we apply a retrieval-augmented large language model (RAG-LLM) to interpret each action in terms of MITRE ATT&CK techniques. The LLM interfaces with a structured knowledge base containing textual descriptions and metadata for all ATT&CK tactics and techniques. For each action, the model retrieves relevant technique entries, compares them to the action's description, and generates an explanation of potential matches. To preserve behavioral continuity, previous actions and inferred techniques are provided as context, allowing the model to reason over evolving attacker strategies (e.g., reconnaissance leading to privilege escalation). The output is a sequence of action-level annotations linking machine telemetry to ATT&CK behavior categories, forming the bridge between raw network events and high-level adversarial reasoning.

## RESULTS

### Evaluation Setup and Comparison Baseline

To assess the accuracy of our Suricata-based inference framework, we benchmark it against results derived using the method in (Hans et al., 2025) that utilizes OPNOTES, which are the real-time journals in which attackers documented their own actions, tools, and objectives as they progressed through the operation. Because these notes were written by the attackers themselves, they provide explicit insight into decision-making, tool choice, and intent. The method therefore captures a semantically rich picture of the operation, which we treat here as a cognitively informed upper bound on interpretive accuracy.

In contrast, the current model receives only machine telemetry, Suricata's encrypted-payload network logs, lacking any linguistic or introspective context. The evaluation thus measures how closely a telemetry-only model can approximate ATT&CK mappings and behavioral timelines inferred from richer human-authored data.

We report the following quantitative dimensions of comparison:
(1) overall **tactic coverage** (presence of each ATT&CK category across participants);
(2) **participant-level precision, recall, and F1** relative to the OPNOTES-derived annotations

This evaluation design allows us to assess not only whether network-based reasoning can recover comparable tactical structure, but also where its information boundaries lie.

### Findings and Comparative Analysis

We begin by evaluating overall detection fidelity between the Suricata-based and OPNOTES-based LLM pipelines using participant-level precision, recall, and F1 scores (Figure 2). Despite the absence of linguistic or cognitive cues, the Suricata-only model achieves high precision across participants, indicating that when it identifies an ATT&CK tactic, it usually aligns with the OPNOTES-derived label. Recall, however, is notably lower and more variable, reflecting the inherent limitation of network telemetry in observing host-centric or intent-driven behaviors. This performance profile suggests that the model reliably captures what is visible on the wire but cannot infer every internal decision recorded in the attacker's own notes.

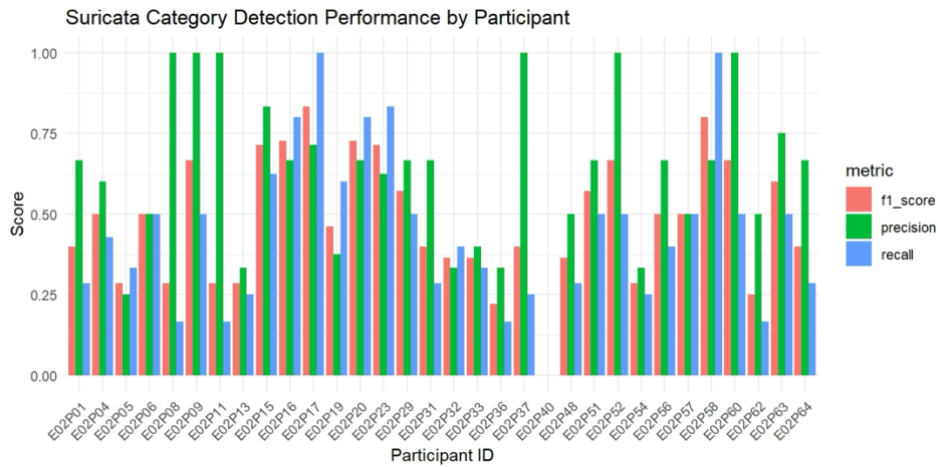

**Figure 2. Suricata Detection Performance compared to OPNOTES-based Detection**

To better understand where these differences arise, we analyze tactic-level coverage and bias across both data sources (Figures 3 and 4). The scatter in Figure 3 shows that Reconnaissance and Lateral Movement fall near the diagonal, indicating strong agreement between systems, whereas Persistence, Command and Control, and Exfiltration are more prevalent in OPNOTES-derived results.

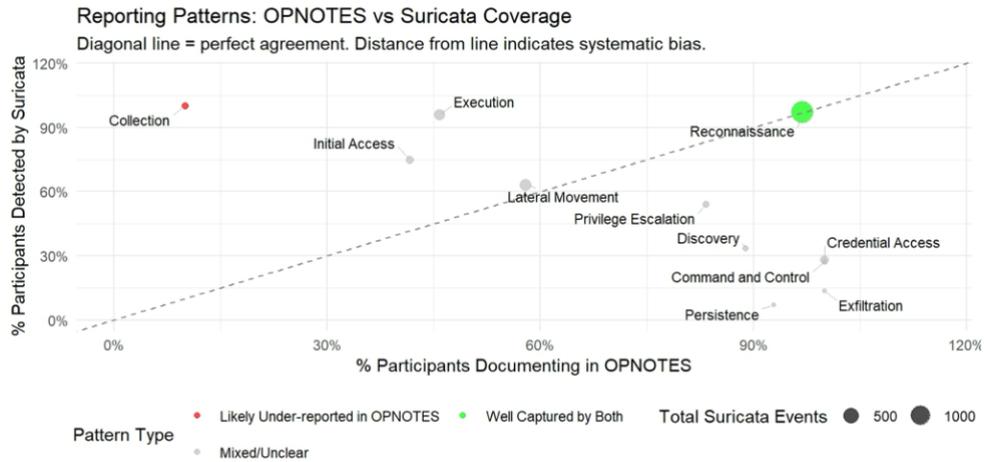

**Figure 3. Per-category detection Suricata vs OPNOTES**

Conversely, Collection behaviors, which are abundant in network traffic but often trivial to attackers, appear more frequently in Suricata-based inference. These trends are quantified in Figure 4, where detection rates exceed 90 % for early, externally visible phases but drop sharply for later or host-bound tactics. Together, these results reveal a consistent pattern: the Suricata model reproduces the structure of the attack where traffic signatures exist, while not doing so well in capturing the cognitive and goal-oriented phases that are absent from telemetry but are often included in the OPNOTES.

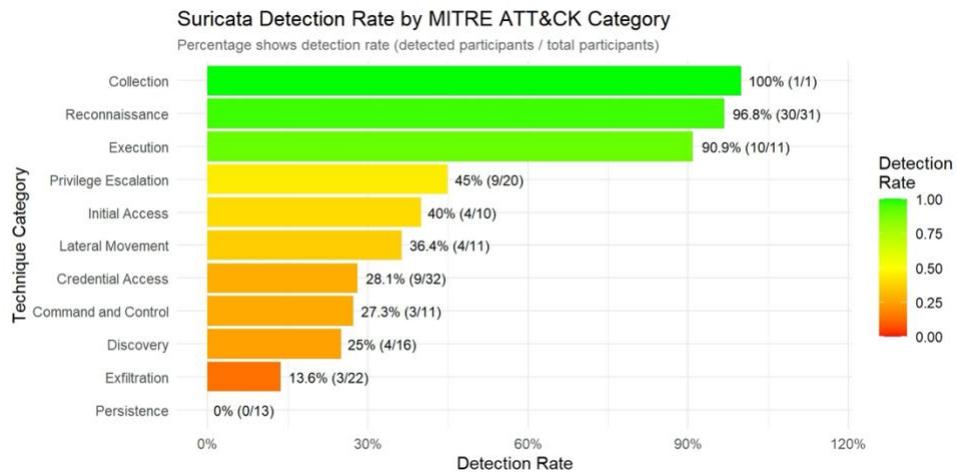

**Figure 4. Per Category Detection Rate**

In summary, the Suricata-based LLM demonstrates that it is indeed possible to infer high-level attacker strategies and cognitive signals directly from network telemetry alone. While its precision remains strong across observable network-layer techniques, performance gaps emerge primarily in categories that depend on information not fully exposed to network sensors—such as credential theft or command execution within hosts. Rather than indicating a failure of reasoning, these gaps highlight the inherent limits of observability in encrypted or host-resident actions. The close alignment between network-derived and OPNOTES-derived technique structures across multiple phases suggests that key elements of attacker intent can still be reconstructed from telemetry, providing a strong foundation for extending cognitive inference to operational settings.

**DISCUSSION**

The results demonstrate that large language models can extract meaningful behavioral structure from low-level network telemetry, approaching the interpretive quality of models that operate on cognitively rich data such as attacker-authored notes. The Suricata-based LLM achieves strong precision and reasonable recall without access to explicit reasoning or unencrypted payloads, indicating that the semantic gap between machine logs and strategic intent can be narrowed through contextual segmentation and ATT&CK-based mapping.

However, the findings also reveal important boundaries. Telemetry-driven inference excels in network-visible phases such as reconnaissance, collection, and execution but struggles to capture host-resident or cognitively motivated behaviors such as persistence or command-and-control adaptation. These patterns reflect a broader principle: **visibility shapes inference**. Where the data record only external manifestations of intent, the model can reconstruct tactics but not necessarily the psychological drivers behind them.

Nevertheless, even partial reconstruction is valuable. By identifying when behavioral shifts occur, such as tool switching, protocol transitions, or pivot patterns, the framework begins to expose latent cognitive signatures embedded in attacker behavior. Such patterns can support higher-level reasoning about traits like loss aversion, risk tolerance, and persistence, extending earlier work on cognitive modeling from human-authored notes to operational telemetry.

**CONCLUSION AND FUTURE WORK**

This work introduces a framework for translating raw intrusion detection logs into interpretable attacker behaviors through hierarchical reasoning with large language models. By segmenting Suricata telemetry into coherent action groups and mapping those actions to MITRE ATT&CK techniques, the system bridges the gap between packet-level events and high-level operational strategy. When benchmarked against an OPNOTES-based LLM baseline, the approach achieves strong alignment in network-visible phases, confirming that much of the structure of adversarial activity can be recovered even without access to human-authored reasoning.

Future work will focus on broadening the framework's scope and deepening its cognitive grounding. One direction is expanding beyond a single network sensor to incorporate multiple sources of data—for example, host telemetry, authentication traces, and contextual system logs—allowing for a more complete reconstruction of attacker operations across layers of observability. Another direction involves connecting this work to concurrent research in real-time bias detection, including efforts to model attacker ambiguity aversion (Carney et al., 2025; Kim et al., 2025) as well as emerging research in the development of bias-aware Theory of Mind architectures capable of anticipating the behavior of boundedly rational adversaries. Finally, deploying the framework in live operational settings would enable continuous inference of attacker strategies and cognitive traits, allowing defensive systems to adapt in real time as adversarial behavior unfolds.

Together, these extensions move toward a new paradigm in cyber defense one that reasons not only about what attackers do, but why they act as they do.

## ACKNOWLEDGMENT

This research is based upon work supported in part by the Office of the Director of National Intelligence (ODNI), Intelligence Advanced Research Projects Activity (IARPA) under Reimagining Security with Cyberpsychology Informed Network Defenses (ReSCIND) program contract N66001-24-C-4504. The views and conclusions contained herein are those of the authors and should not be interpreted as necessarily representing the official policies, either expressed or implied, of ODNI, IARPA, or the U.S. Government. The U.S. Government is authorized to reproduce and distribute reprints for governmental purposes notwithstanding any copyright annotation therein.